\documentclass[reprint,amsmath,amssymb,aps,prc,superscriptaddress,floatfix,nofootinbib]{revtex4-2}
\usepackage[T1]{fontenc}
\usepackage{graphicx}
\usepackage{dcolumn}
\usepackage{amsmath}
\usepackage{bm}
\usepackage{hyperref}
\usepackage{epstopdf}

\begin{document}

\preprint{APS/123-QED}

\title{Effective Color Dipole Approach to Color Transparency in $\rho^0$ Electroproduction}

\author{Tae Keun Choi}
\email{tkchoi@yonsei.ac.kr} \affiliation{Department of Physics and
Engineering Physics, Yonsei University, Wonju 26493, Korea}

\author{Kook-Jin Kong}
\email{kong@kau.ac.kr} \affiliation{Research Institute of Basic
Science, Korea Aerospace University, Goyang 10540, Korea}

\author{Byung-Geel Yu}
\email{bgyu@kau.ac.kr}
\affiliation{Research Institute of Basic Science, Korea Aerospace
University, Goyang 10540, Korea}
\affiliation{Center for Exotic Nuclear
Studies, Institute for Basic Science, Daejeon 34126, Korea}

\date{\today}

\begin{abstract}
We investigate nuclear transparency in exclusive $\rho^0$ electroproduction
on $^{12}$C and $^{56}$Fe nuclei within a multi-channel final-state
interaction (FSI) framework that explicitly incorporates the kinematic
decay length effect (DLE) arising from the short-lived
$\rho^0\to\pi^+\pi^-$ decay. The purely kinematic and nuclear mechanisms
prove insufficient to account for the CLAS data: the DLE alone cannot
generate the observed $Q^2$-dependent enhancement, and the inclusion of
nuclear shadowing further deepens the disagreement, so that a
compensating reduction of the in-medium attenuation---the hallmark of
color transparency (CT)---is required. To incorporate the color dynamics
of the initially compact $q\bar{q}$ configuration, we replace the
empirical Quantum Diffusion Model (QDM) ansatz for the initial
interaction cross section $\sigma_h(Q^2)$ of the point-like configuration
(PLC) by an effective Color Dipole Model (CDM) boundary condition,
evaluated through a normalized dipole-weighted $\gamma^*\to\rho^0$
transition overlap. Combined with the standard linear QDM transport at an
effective in-medium expansion scale $\Delta m^2 = 0.3$~GeV$^2$, the CDM
boundary condition reproduces both the magnitude and the $Q^2$ dependence
of the data for both targets. A $\chi^2$ analysis quantifies the
pronounced separation between the non-CT and CT-based descriptions and
thereby supports the onset of color transparency in the $\rho^0$ channel
beyond what kinematic decay-length effects can accommodate.
\end{abstract}

\pacs{24.85.+p, 25.30.Rw, 13.60.Le, 12.38.Aw}

\maketitle

\section{Introduction}

The phenomenon of Color Transparency (CT), a signature prediction
of perturbative Quantum Chromodynamics (QCD) in high-energy
exclusive nuclear processes, asserts that a hadron can traverse
the nuclear medium with reduced attenuation if it is produced in a
spatially compact state known as a Point-Like Configuration
(PLC)~\cite{Mueller1982,Brodsky1982}. Because the color dipole
moment of a small color-singlet system vanishes with its
transverse size, its strong interaction with surrounding nucleons
is substantially suppressed---a feature known as color
screening~\cite{Nikolaev1991,Golec1998,Iancu2004}. Tracking the
evolution of this compact system into a fully formed hadron as it
propagates through the nucleus provides a unique window into the
transition from non-perturbative hadronic degrees of freedom to
the fundamental domain of quarks and gluons.

Extensive experimental efforts have sought to establish the onset
of CT across various kinematic domains. While clear evidence of CT
has been observed in diffractive pion dissociation into di-jets at
Fermilab~\cite{Aitala2001} and in exclusive pion electroproduction
at Jefferson Lab (JLab)~\cite{Clasie2007,XQian2010}, the
situation in the baryon sector remains less clear. Recent
high-precision data for the $^{12}\text{C}(e,e'p)$ reaction from
JLab Hall C showed no indication of CT up to $Q^2 = 14.2\text{
GeV}^2$~\cite{Bhetuwal2021}, stimulating renewed theoretical
discussions regarding the different expansion timescales and
configuration dynamics between mesons and baryons.

In the meson sector, exclusive $\rho^0$ electroproduction on
nuclear targets, $A(e,e'\rho^0)A'$, has served as a principal 
testing ground for CT dynamics. The JLab CLAS collaboration reported a
noticeable increase in nuclear transparency with increasing
virtuality $Q^2$ for both $^{12}\text{C}$ and $^{56}\text{Fe}$
targets~\cite{ElFassi2012}. However, interpreting these data is
complicated by the presence of competing kinematic mechanisms,
most notably the Decay Length Effect (DLE)~\cite{Frankfurt2008,
Gallmeister2011}. The DLE arises because the very short lifetime
of the $\rho^0$ meson allows it to decay into a $\pi^+\pi^-$ pair
inside the nuclear volume, leading to a momentum-dependent
multi-channel absorption. At lower energies, this decay length can
be comparable to or smaller than the nuclear radius, leading to a
natural variation in the transparency ratio with energy that can
mimic the signature of CT.

Several theoretical models have been proposed to analyze the CLAS
data. The Kopeliovich-Nemchik-Schmidt (KNS) light-cone dipole
approach~\cite{Kopeliovich2002} and the Frankfurt-Miller-Strikman
(FMS) semi-classical transport model~\cite{Frankfurt2008} have
attempted to balance the geometric and quantum aspects of the
reaction. Nonetheless, drawing a definitive conclusion has been
hindered by model-dependent uncertainties, such as the precise
modeling of hadronic wave functions and the phenomenological
parametrization of the PLC expansion rate.

In this work, we address these challenges by formulating a hybrid theoretical
framework that incorporates a Color Dipole Model (CDM)~\cite{Nikolaev1991,Golec1998} inspired initial
boundary condition into a semi-classical multi-channel final-state
interaction (FSI) transport description. Within the conventional
Quantum Diffusion Model (QDM), a point-like configuration (PLC) is
assumed to expand linearly toward its fully formed hadronic state
while propagating through the nuclear medium. However, the initial
PLC interaction cross section in the standard QDM is determined through
a constituent-quark power-law ansatz rather than from the transverse-size
distribution selected at the production vertex. The central novelty of the
present approach is not a replacement of the QDM transport picture,
but a physically motivated determination of the initial PLC interaction
cross section $\sigma_{\text{h}}(Q^2)$ via a dipole-weighted overlap of
the $\gamma^*$ and $\rho^0$ light-cone wave functions.

The present analysis differs from previous studies in
several respects. First, we implement the deuteron wave function
obtained from the Paris potential, incorporating the
short-range repulsive core and tensor correlations.
This provides a realistic treatment of the deuteron
structure and a reliable normalization for the
transparency ratio $T_A/T_D$.  Second, we explicitly convolve the
exponential $\rho^0$ decay probability with the FSI integrations,
so that the DLE is treated as a geometric kinematic effect rather
than a correction factor.
Third, we demonstrate that an accurate treatment of
the DLE is essential: neglecting the DLE leads to a
substantial overestimation of the measured transparency,
whereas retaining the DLE and nuclear shadowing alone
still fails to reproduce the observed $Q^2$ dependence.
Fourth, we determine the effective in-medium expansion scale
$\Delta m^2$ within the Effective CDM framework through a $\chi^2$
comparison with the CLAS data and subsequently use the same value in
the conventional QDM calculation without readjustment, which allows
the non-CT and CT-based descriptions to be discriminated on a common
quantitative footing.

\section{Theoretical Framework}

The exclusive electroproduction of $\rho^0$ mesons on a nuclear
target $A$ is induced by a virtual photon $\gamma^*$ with
four-momentum $q = (k - k')$, where $k$ and $k'$ are the
four-momenta of the incident and scattered electrons,
respectively. The virtuality of the photon is defined as $Q^2 =
-q^2$. The nuclear transparency $T_A$ for exclusive $\rho^0$
electroproduction is defined as the ratio of the nuclear cross
section to the free nucleon cross section scaled by the atomic
mass number $A$. Within the semi-classical Glauber framework, it
is expressed as an integral over the impact parameter $\mathbf{b}$
and the longitudinal interaction vertex $z$:

\begin{eqnarray}
T_A = \frac{1}{A} \int d^2\mathbf{b}
\int_{-\infty}^{\infty} dz \, \rho(\mathbf{b}, z)
S_{\text{ISI}}(\mathbf{b}, z) S_{\text{FSI}}(\mathbf{b}, z).
\label{eq:TA_general}
\end{eqnarray}

Here, \(\rho(\mathbf{b}, z)\) denotes the nuclear density
distribution normalized to \(A\). $S_{\text{FSI}}(\mathbf{b}, z)$
represents the survival probability of the propagating state
against final-state interactions (FSI) as it exits the nuclear
medium from the production point $(\mathbf{b}, z)$, and the factor
$S_{\text{ISI}}(\mathbf{b}, z)$ accounts for the initial-state
interactions (ISI) of the incident virtual photon propagating
through the nuclear medium before the hard scattering vertex.
$S_{\text{ISI}}$ plays a role in incorporating nuclear shadowing
effects, and its formulation closely follows the theoretical
framework discussed in Ref.~\cite{Kong2025}. The initial state
shadowing is fixed at $\sigma_{\text{ISI}} = 25\,\text{mb}$.
This value corresponds approximately to the effective
vector-meson–nucleon interaction strength employed in Ref.~\cite{Kong2025}.
To mitigate arbitrary normalization uncertainties, the transparency
ratio $T_A/T_D$ is constructed by evaluating the deuteron
transparency $T_D$ using the same convolution framework.

\subsection{Nuclear Density Profiles}
We utilize a harmonic oscillator (HO) distribution for
$^{12}\text{C}$ and a two-parameter Fermi (2pF) profile for
$^{56}\text{Fe}$ based on standard electron scattering data.
For $^{12}\text{C}$, the HO density
parameters are characterized by a size parameter $a = 1.692 \text{
fm}$ and a fractional parameter $\alpha = 1.082$ \cite{DeJager1974}. For the heavier
$^{56}\text{Fe}$ nucleus, the 2pF profile employs a half-density
radius $R = 4.111 \text{ fm}$ and a surface diffuseness parameter
$a_0 = 0.558 \text{ fm}$ \cite{DeVries1987}.

For the deuteron ($A=2$), we employ the Paris potential
wave function~\cite{Lacombe1981}, which incorporates
the short-range repulsive core and tensor correlations
inherent in the deuteron structure. Compared with simplified 
deuteron density profiles, the Paris
wave function modifies the spatial weighting of short-distance
configurations and leads, within the present convolution framework,
to a reduced deuteron transparency $T_D$.
Since the nuclear transparency is extracted as the ratio
$T_A/T_D$, the treatment of the deuteron wave function
has a direct quantitative impact on the comparison with
experimental data. This sensitivity highlights the
importance of a realistic description of the deuteron
reference state.

The Paris wave function is evaluated numerically using the
parameterization of Ref.~\cite{Lacombe1981}. The resulting
deuteron density is normalized according to
$\int d^3r\,\rho_d(r)=2,$
and the numerical normalization is verified to better than
$0.1\%$.

\subsection{FSI Convolution and the QDM Expansion}
The FSI treats the decay position $z_d$ as an explicit convolution
integral over the exponential decay probability:
\begin{eqnarray}
\label{eq:strict_fsi}
&&S_{\text{FSI}}(b, z) = \int_{z}^{\infty} dz_d \left( \frac{1}{l_d} e^{-(z_d - z)/l_d} \right) \nonumber \\
&&\times \exp\left[ - \int_{z}^{z_d} \sigma_{\text{eff}}(x, Q^2)
\rho(x) dx - \sigma_{\pi\pi}\int_{z_d}^{\infty} \rho(y) dy
\right], \quad \quad
\end{eqnarray}
where $x$ and $y$ denote the longitudinal paths of the $\rho^0$
and the pion pair, respectively. $l_d$ is the decay length for
$\rho^{0}\rightarrow\pi^{+}\pi^{-}$. The dynamic decay length in
the laboratory frame is formulated as:
\begin{equation}
    l_d = \gamma v \tau = \left( \frac{p_\rho}{m_\rho} \right) \left( \frac{\hbar c}{\Gamma_\rho} \right).
\end{equation}
The decay length $l_d$ is computed dynamically for each $Q^2$ bin,
ensuring a precise geometric representation of the decay
kinematics. Within the QDM framework, the PLC expands linearly
along its formation length $l_f = 2p_\rho\hbar c / \Delta m^2$,
where $\Delta m^2$ represents the effective expansion scale. The
dynamically evolving cross-section is written as:
\begin{equation}
\label{eq:sigma_eff_qdm} 
\sigma_{\rm eff}(z,Q^2)=
\begin{cases}
\sigma_h(Q^2)
+\left[\sigma_{\rho N}-\sigma_h(Q^2)\right]
\dfrac{z}{l_f},
& z\le l_f,\\[2mm]
\sigma_{\rho N},
& z>l_f.
\end{cases}
\end{equation}
Here, the initial effective interaction cross section of the PLC
is conventionally estimated from the constituent-quark transverse
momentum $\langle k_t^2 \rangle^{1/2}\simeq 0.35$
GeV$/c$ \cite{Farrar1988}, yielding
\begin{equation}
\label{eq:qdm_sigma_h}
\sigma_{\text{h}}(Q^2) = \sigma_{\rho N} \frac{ \langle
4k_t^2\rangle}{Q^2}.
\end{equation}
The fully expanded $\rho$-$N$ cross-section is fixed at
$\sigma_{\rho N} = 2.5\,\text{fm}^2$ ($25$ mb), and the
effective pion-pair absorption cross section is set to $\sigma_{\pi\pi} =
5.0\,\text{fm}^2$ ($50$ mb).

To isolate the role of DLE, additional calculations are performed with the DLE
switched off. In this case, the pion-pair interaction
is removed by setting $\sigma_{\pi\pi}=0$, and the
decay length is fixed to $l_d=40$ fm. Since this
length substantially exceeds the diameter of even the
heaviest nuclei considered here ($2R_A\lesssim20$ fm),
the produced $\rho^0$ meson effectively decays
outside the nucleus, thereby eliminating pion
final-state interactions.

\subsection{Effective CDM Boundary Condition}
In the traditional QDM, the cross section expands linearly,
but the initial PLC effective interaction cross section
$\sigma_{\text{h}}$ is assigned by the purely empirical
constituent-quark power-law ansatz of Eq.~\eqref{eq:qdm_sigma_h},
which lacks a direct connection to the $q\bar{q}$ size
distribution actually selected at the production vertex.
To establish a more physically grounded initial condition
while preserving the geometric transport structure of the
linear expansion, we construct a hybrid framework. 
Rather than assigning $\sigma_h(Q^2)$ through the empirical
power-law prescription of Eq.~(5), we determine it from the
CDM-inspired overlap construction at the production vertex.
The evaluated $\sigma_{\text{h}}(Q^2)$ is then substituted directly
into Eq.~\eqref{eq:sigma_eff_qdm} to govern the FSI expansion.
This procedure is motivated by the separation of dynamical scales
in the reaction: the production vertex is governed by the
transverse-size distribution of the compact $q\bar q$ configuration,
whereas the subsequent in-medium evolution is adequately described
by the established semiclassical QDM transport framework.

The Effective CDM is introduced here only at the level
of the initial condition.
The QDM linear expansion law
and all subsequent FSI integrations are unchanged; only the
starting value $\sigma_{\text{h}}(Q^2)$ is replaced by the
CDM overlap integral defined below. Within this
interpretation, the fitted $\Delta m^2$ characterizes the
effective in-medium expansion rate of the initially compact
$\rho^0$ configuration rather than a literal single-state
excitation mass gap.

The effective PLC interaction cross section $\sigma_h(Q^2)$ is
defined through a normalized dipole-weighted transition overlap:
\begin{equation}
\sigma_h(Q^2)
=
\int d^2r
\int_0^1 dz\,
P_{\gamma^*\to\rho}(r,z;Q^2)
\hat{\sigma}_{\rm dip}^{\rm eff}(r).
\label{eq:cdm}
\end{equation}
Here $r$ denotes the transverse separation of the $q\bar q$
dipole and $z$ is the light-cone momentum fraction carried by
the quark. The transition-overlap weight is defined by
\begin{equation}
P_{\gamma^*\to\rho}(r,z;Q^2)
=
\frac{\Psi_\rho^*(r,z)\Psi_{\gamma^*}(r,z;Q^2)}
{\displaystyle
\int d^2r\int_0^1 dz\,
\Psi_\rho^*(r,z)\Psi_{\gamma^*}(r,z;Q^2)},
\label{eq:overlap}
\end{equation}
and satisfies the normalization condition
\begin{equation}
\int d^2r\int_0^1 dz\,
P_{\gamma^*\to\rho}(r,z;Q^2)=1.
\end{equation}
Accordingly, $P_{\gamma^*\to\rho}$ provides a normalized
transition-overlap weight that characterizes the transverse
dipole configurations selected in the $\gamma^*\to\rho^0$
production process. It is introduced here as a phenomenological
measure of transverse-size selection and should not be interpreted
as a genuine quantum-mechanical probability density.
Within this construction, Eq.~\eqref{eq:cdm} defines an effective
interaction cross section associated with the configurations selected
at the production vertex, rather than a strict quantum-mechanical
expectation value.

The denominator in Eq.~\eqref{eq:overlap} removes the overall transition-overlap normalization
and isolates the $Q^2$-dependent transverse-size filtering effect. In the small-$r$
region relevant at large $Q^2$, the virtual-photon wave function behaves
approximately as $\Psi_{\gamma^*}\propto QK_0(\epsilon r)$, with $\epsilon\sim Q$,
so that the overlap is dominated by dipole sizes $r\sim1/Q$.
Since $\hat{\sigma}_{\rm dip}(r)\propto r^2$ for small dipoles, the numerator
contains an additional factor $r^2\sim Q^{-2}$ relative to the denominator.
Parametrically, the denominator behaves approximately
as $Q^{-1}$, whereas the numerator acquires an additional
suppression factor of $r^2\sim Q^{-2}$, leading to an
overall behavior close to $\sigma_h(Q^2)\propto Q^{-2}$.
This behavior is consistent with the power-law scaling
of the standard QDM ansatz, Eq.~\eqref{eq:qdm_sigma_h}.

For longitudinal polarization, the light-cone wave function for
the virtual photon is given by \cite{Dosch1997}:
\begin{equation}
\Psi_{\gamma^*_L}(\mathbf{r}, z; Q^2) = \sqrt{N_c} \frac{e \,
\hat{e}_{\rho}}{2\pi} \cdot 2z(1-z) Q \, K_0(\epsilon r),
\end{equation}
where $N_c = 3$, $\hat{e}_{\rho} = 1/\sqrt{2}$,
and $\epsilon^2 =
z(1-z)Q^2 + m_q^2$ with $m_q = 0.14$ GeV. For the $\rho^0$ meson, 
we adopt a simplified Gaussian spatial
profile motivated by the Boosted Gaussian (BG) framework
\cite{Nemchik1994,Nemchik1996}:
\begin{equation}
\Psi_{\rho_L}(\mathbf{r},z)
=
N_L z(1-z)
\exp\left(-\frac{r^2}{2R_\rho^2}\right),
\end{equation}
where $R_\rho=0.73$ fm is chosen as the vacuum electromagnetic
radius and $N_L$ is the normalization constant. This simplified
form retains the transverse-size filtering relevant to the present
construction while avoiding the introduction of additional
model-dependent parameters.

As a robustness check, we have repeated the calculation using
the full Boosted Gaussian form, including its $z$-dependent
transverse width and endpoint-suppression structure. The resulting
$\sigma_h(Q^2)$ differs from that obtained with the simplified
profile by approximately $8$--$9\%$, while the corresponding
change in the nuclear transparency ratio $T_A/T_D$ remains below
$0.5\%$ over the kinematic range considered here. Thus, although
the detailed initial effective cross section exhibits a moderate
wave-function dependence, its impact on the final transparency
observable is negligible within the accuracy of the present
analysis.

For the dipole-nucleon interaction, we employ a
Golec-Biernat--W\"usthoff (GBW)-inspired effective dipole
cross section~\cite{Golec1998,Golec1999}:
\begin{equation}
\hat{\sigma}_{\rm dip}^{\rm eff}(r)
=
\sigma_0
\left[
1-\exp\left(
-\frac{r^2}{4R_{\rm eff}^2}
\right)
\right],
\label{eq:sigma_dip_eff}
\end{equation}
where $R_{\rm eff}$ is not interpreted as the saturation radius
of the original small-$x$ GBW model, but rather as an effective
transverse scale characterizing the onset of color screening in
the present intermediate-energy kinematics. In the original GBW
parameterization, the transverse scale $R_{\rm eff}$ is $x$ dependent,
i.e., $R_0(x)=Q_0^{-1}(x/x_0)^{\lambda/2}$. The Bjorken variable 
$x=\frac{Q^2+M_\rho^2}{W^2+Q^2-m_N^2}$
lies approximately in the range $x\simeq0.1$--$0.3$
for the CLAS kinematics considered here. Since this region is
outside the small-$x$ domain in which the original GBW model was
constrained, we do not employ its explicit $x$ dependence. Instead,
we fix $R_{\rm eff}=0.5$ fm, corresponding to a crossover scale
$r\sim2R_{\rm eff}\sim1$ fm of typical hadronic size. The
normalization $\sigma_0=2.3~\mathrm{fm}^2$ is retained as a
representative hadronic interaction scale consistent with the
original GBW fit~\cite{Golec1998}, without further optimization.

\section{Results and Discussion}

All calculations are performed in the laboratory
frame using kinematic conditions corresponding
to the CLAS experiment of Ref.~\cite{ElFassi2012}.
The transparency is evaluated
using an approximate effective average invariant mass $W = 2.2$
GeV (representing the CLAS experimental range of $W \approx 2.0 -
2.4$ GeV) and momentum transfer in the forward direction ($t
\approx t_{min}$, bounded by $-t \le 0.4$ GeV$^2$).

\begin{figure}[t]
\centering
\includegraphics[width=0.45\textwidth]{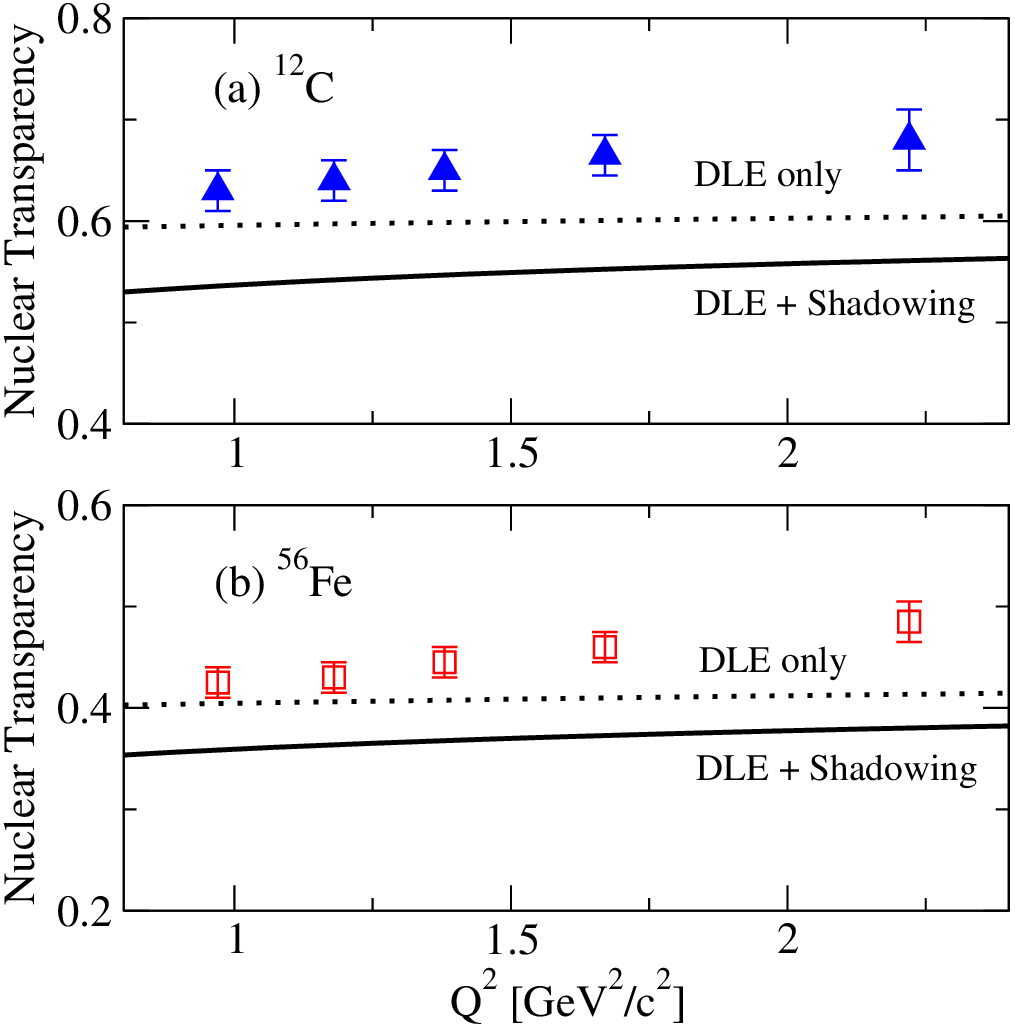}
\caption{Nuclear transparency ratio $T_A/T_D$ for $^{12}$C and
$^{56}$Fe. Upward triangles and squares represent data from the
JLab CLAS Collaboration \cite{ElFassi2012}. 
The dotted line represents the Glauber calculation including only
the kinematic Decay Length Effect (DLE), with the deuteron reference
state described by the Paris wave function~\cite{Lacombe1981}.
The solid line displays the
calculation where both the DLE and standard nuclear
shadowing effects are included. } \label{fig1}
\end{figure}

In Fig.~\ref{fig1}, we examine the capability of purely
kinematic and standard nuclear mechanisms to account
for the observations. The dotted line displays the
transparency obtained when only the kinematic DLE
is included, while the solid line additionally includes
standard nuclear shadowing effects.
Because the CLAS observable is defined as the ratio
$T_A/T_D$, the extracted transparency depends on the
treatment of the deuteron reference state. The realistic
Paris wave function therefore provides a reliable
normalization for the transparency ratio.

As $Q^2$ increases, the relativistic decay length
$l_d$ increases, reducing the probability of pion
absorption inside the nucleus and generating a weak
upward trend in both curves. However,
this kinematic rise alone is insufficient to explain
the steep $Q^2$-dependent slope of the CLAS data.
Furthermore, the persistent discrepancy suggests that purely
geometric mechanisms, including the DLE and
standard nuclear shadowing, are insufficient and that
an additional dynamical mechanism associated with
the reduced interaction strength of compact PLC
configurations is required.

\begin{figure}[t]
\centering
\includegraphics[width=0.45\textwidth]{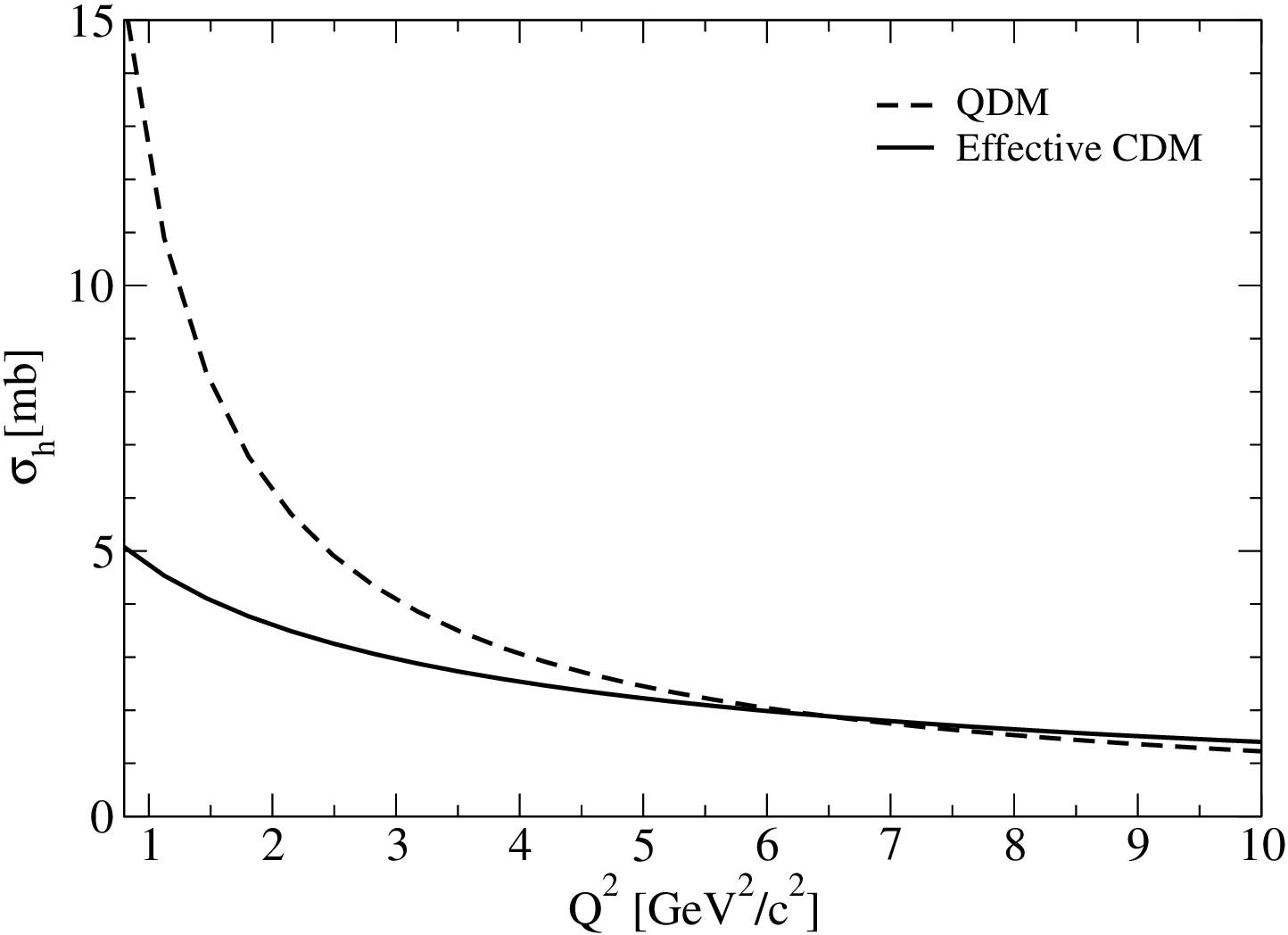}
\caption{Effective PLC interaction cross section,
$\sigma_h(Q^2)$, as a function of $Q^2$.
The dashed curve corresponds to the conventional
QDM parametrization of Eq.~\eqref{eq:qdm_sigma_h}, while the solid
curve shows the dipole-weighted effective interaction
cross section obtained from the Effective CDM
overlap integral of Eq.~\eqref{eq:cdm}. The Effective CDM
yields a reduced interaction cross section,
particularly at low and intermediate $Q^2$,
reflecting the preferential selection of compact
transverse $q\bar q$ configurations in the
$\gamma^*\rightarrow\rho^0$ transition.
} \label{fig2}
\end{figure}

Before presenting the nuclear transparency results,
it is instructive to examine the effective PLC interaction
cross section itself. Figure~\ref{fig2} shows the $Q^2$ dependence
of $\sigma_h(Q^2)$ obtained from the dipole-weighted overlap
integral of Eq.~\eqref{eq:cdm}, together with the conventional QDM
ansatz of Eq.~\eqref{eq:qdm_sigma_h}. The Effective CDM yields a
smaller interaction cross section, particularly in the low- and
intermediate-$Q^2$ region relevant to the present CLAS kinematics.
This reduction weakens the initial attenuation of the PLC during
its propagation through the nuclear medium, thereby enhancing the
survival probability of compact configurations.

The physical origin of this difference lies in the distinct
specification of the initial PLC boundary condition.
The conventional QDM imposes a prescribed
$1/Q^2$ dependence on the initial PLC interaction cross section,
whereas the Effective CDM determines $\sigma_h(Q^2)$ from the
transverse-size distribution selected by the production overlap.
As shown in Fig.~\ref{fig2}, the latter yields a substantially
smaller initial interaction cross section over the measured
kinematic range while remaining below the fully expanded
$\rho N$ cross section. This provides a physically consistent
PLC boundary condition without altering the subsequent QDM
expansion dynamics.

Consequently, the Effective CDM predicts a larger nuclear
transparency than the standard QDM. This enhancement reflects the
reduced interaction strength associated with the compact transverse
configurations selected by the production overlap. Furthermore, the
$Q^2$-dependent decrease of $\sigma_h(Q^2)$ contributes to the rising
$Q^2$ dependence of the calculated transparency. As seen in
Fig.~\ref{fig2}, the numerical result approaches an approximate
$Q^{-2}$ behavior, consistent with the
asymptotic scaling argument following Eq.~\eqref{eq:cdm}. The extended
$Q^2$ range shown in Fig.~\ref{fig2} is included to illustrate this
asymptotic trend beyond the CLAS kinematic region.

\begin{figure}[tbp]
\centering
\includegraphics[width=0.45\textwidth]{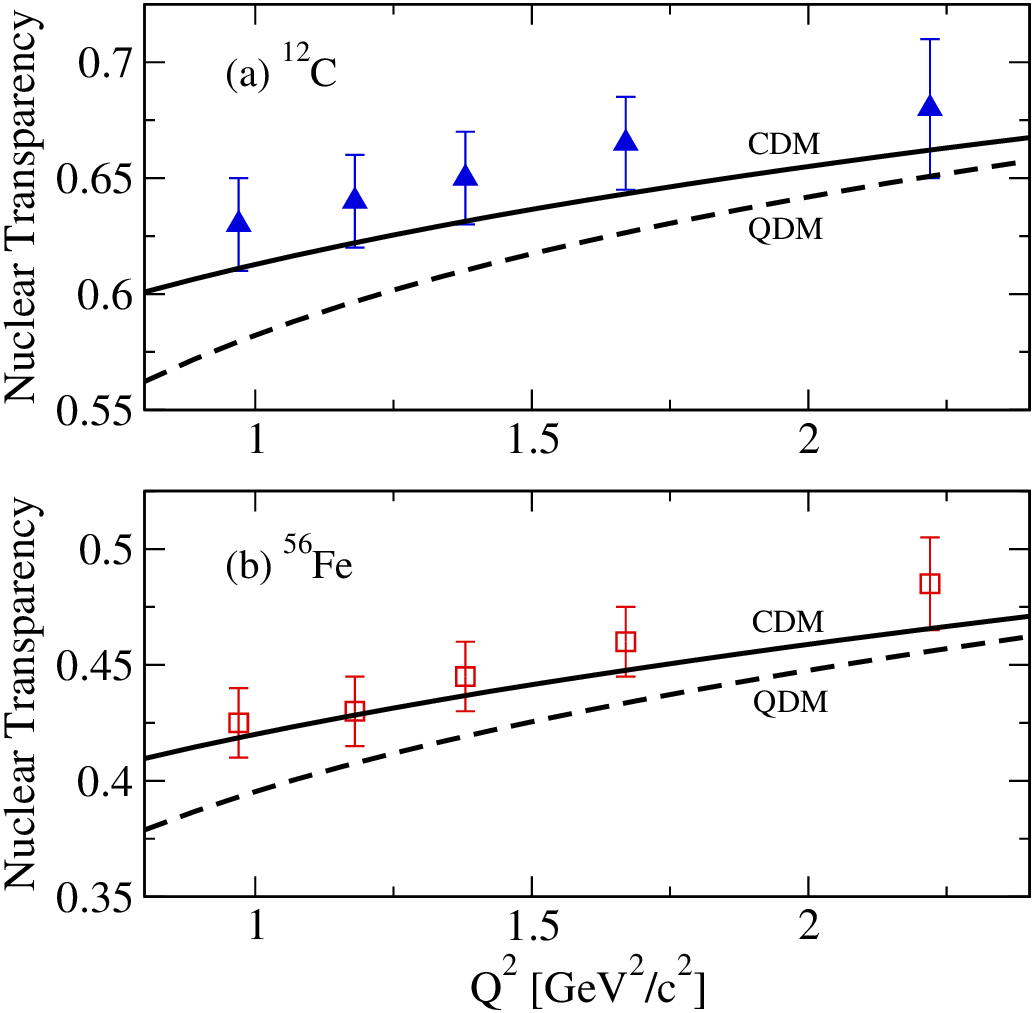}
\caption{Nuclear transparency of exclusive $\rho^0$
electroproduction on $^{12}$C and $^{56}$Fe as a function of $Q^2$
within the hybrid Effective CDM framework.
The solid lines represent predictions obtained with the Effective
CDM boundary condition, while the dashed lines correspond to the standard
QDM prescription. For the combined data sets, the Effective CDM
provides an improved quantitative description of the
data compared with the standard QDM.
} \label{fig3}
\end{figure}

To elucidate the role of the CDM boundary condition, Figure~\ref{fig3}
displays the calculated transparencies for $^{12}$C and $^{56}$Fe
as a function of $Q^{2}$. The solid lines represent the Effective CDM
results and the dashed lines the standard QDM at the same
expansion scale, both compared with the CLAS data.
The underlying mechanism is straightforward. The CDM boundary condition
$\sigma_{\text{h}}(Q^2)$ evaluated via Eq.~\eqref{eq:cdm} is
governed by the transverse size of the $q\bar{q}$ configuration
selected at the production vertex. As $Q^2$ increases, the virtual
photon wave function $\Psi_{\gamma^*_L}\propto K_0(\epsilon r)$
concentrates weight at smaller transverse separations $r$. Since
$\hat{\sigma}_{\text{dip}}\propto r^2$ at small $r$, the
overlap integral yields a smaller $\sigma_{\text{h}}(Q^2)$ at
higher $Q^2$,  leading to a reduced in-medium attenuation
during the early stage of propagation. 
This overlap-driven modification of the initial attenuation
distinguishes the Effective CDM boundary condition from the
prescribed $1/Q^2$ scaling of the conventional QDM ansatz and
provides the additional $Q^2$ dependence required to reproduce the observed slope.

The value $\Delta m^2=0.3~\mathrm{GeV}^2$ is obtained within
the Effective CDM framework and is subsequently used, without
readjustment, in the conventional QDM calculation to isolate the
effect of the initial PLC boundary condition.

A $\chi^2$ comparison with ten CLAS data points for both targets is
performed using the effective expansion scale $\Delta m^2=0.3$ GeV$^2$.
The four calculations fall into two categories: the non-CT descriptions
(DLE only and DLE with shadowing), in which the propagating $\rho^0$
interacts with the full hadronic strength, and the CT-based descriptions
(QDM and Effective CDM), in which the initially compact configuration
interacts with a reduced cross section that evolves toward
$\sigma_{\rho N}$.
In Fig. \ref{fig3} the Effective CDM yields $\chi^2/N=0.62$ 
for the combined data set, whereas the
standard QDM, evaluated with the same expansion scale without
readjustment, gives $\chi^2/N=3.45$.
In contrast, the $\chi^2$ comparison for the curves in Fig. \ref{fig1} yields $\chi^2/N = 6.56$ for the DLE-only calculation and $\chi^2/N = 24.7$ when shadowing is included.
Note that including the physically required shadowing worsens the agreement, indicating that a compensating reduction of the in-medium attenuation
---the defining signature of CT---is indispensable rather than optional.


The substantially smaller $\chi^2$ obtained with the
Effective CDM indicates that a $Q^2$-dependent dipole
boundary condition provides a markedly improved
description of the data. While the conventional QDM
captures the overall increasing trend with $Q^2$, the
Effective CDM reproduces both the magnitude and the
$Q^2$ dependence of the transparency significantly more
accurately. This quantitative improvement is achieved
without modifying the subsequent QDM transport
dynamics, suggesting that the initial PLC boundary
condition plays a major role in the description of the
present data.

The fitted value $\Delta m^2 = 0.3~\mathrm{GeV}^2$ should be
interpreted as an effective in-medium expansion scale within
the present semi-classical transport framework rather than as
a direct identification with a specific $\rho'$ excitation.
The extracted value is smaller than those typically adopted
in conventional QDM analyses. However, this difference does
not imply a smaller physical excitation mass gap.
Because the present framework already incorporates a reduced
initial PLC interaction through the CDM boundary condition,
the fitted expansion scale effectively absorbs residual medium
and transport effects. Consequently, the extracted $\Delta m^2$
cannot be directly compared with values obtained in standard
QDM implementations.

\begin{figure}[tbp]
\centering
\includegraphics[width=0.45\textwidth]{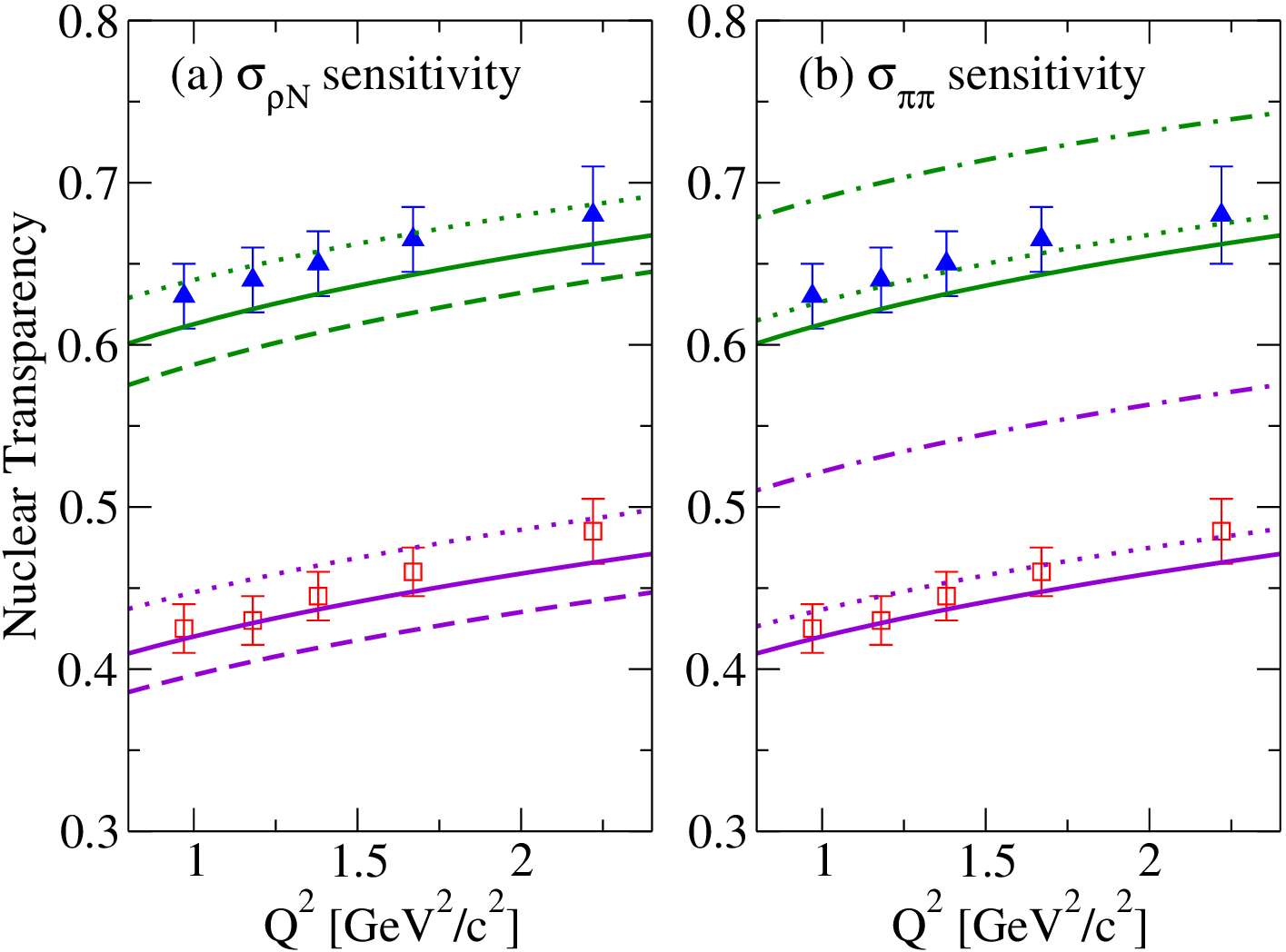}
\caption{Sensitivity of the calculated nuclear transparency within
the hybrid framework at $\Delta m^2 = 0.3~\mathrm{GeV}^2$.
(a) Dependence on the $\rho N$ total cross section,
with $\sigma_{\rho N}=20$, 25, and 30 mb represented by
the dotted, solid, and dashed curves, respectively.
(b) Dependence on the effective pion-pair absorption
cross section, where $\sigma_{\pi\pi}=40$ and $50$ mb are
shown by the dotted and solid curves, respectively.
The data points are the CLAS measurements.
The dash-dotted curve shown in the right panel
corresponds to the calculation with the DLE deactivated.
} \label{fig4} \end{figure}

Figure~\ref{fig4} examines the sensitivity of the calculated
nuclear transparency to variations of the effective hadronic
cross sections, evaluated at the fitted expansion scale
$\Delta m^{2}=0.3~\mathrm{GeV}^{2}$. The left panel shows the
dependence on the $\rho^{0}$-nucleon total cross section
$\sigma_{\rho N}$. The transparency exhibits a noticeable
sensitivity to $\sigma_{\rho N}$, reflecting its direct role in
the attenuation of the propagating $\rho^0$ configuration.

The right panel shows the corresponding dependence on the effective
pion-pair absorption cross section $\sigma_{\pi\pi}$. Variations in
$\sigma_{\pi\pi}$ produce only a modest shift in the overall
transparency magnitude without significantly altering its
$Q^2$ dependence. This relatively weak sensitivity arises because
the Lorentz boost increases the decay length $l_d$ with increasing
$Q^2$, causing an increasing fraction of the $\rho^0$ mesons to
decay outside the densest nuclear region and thereby reducing the
influence of pion final-state interactions.

To cleanly separate the genuine CT effect from this
kinematic background, we also plot the transparency
trajectory with the DLE deactivated (dash-dot curve in the right panel).
Without the DLE, the theoretical calculations drastically
overestimate in the low-$Q^2$ region. This indicates that
while the high-$Q^2$ enhancement originates from dynamical color expansion,
a proper treatment of the DLE remains a
critical component in establishing the experimental data.

\section{Summary and Conclusion}

We have studied nuclear transparency in exclusive $\rho^0$
electroproduction on $^{12}$C and $^{56}$Fe within a multi-channel FSI
framework that combines an explicit convolution of the
$\rho^0\to\pi^+\pi^-$ decay probability with the FSI propagation, a
realistic deuteron reference state given by the Paris wave function, and
a CDM-inspired boundary condition for the initial PLC interaction cross
section in place of the empirical QDM power-law ansatz.

The central result is a quantitative separation between the non-CT and
CT-based descriptions of the CLAS data. The purely kinematic and nuclear
mechanisms are incompatible with the observed $Q^2$ dependence: the DLE
alone gives $\chi^2/N = 6.56$, and adding the physically required
shadowing worsens the description to $\chi^2/N = 24.7$, 
so that a compensating reduction of the in-medium attenuation is
indispensable, which constitutes evidence for the onset of CT.
Among the CT-based descriptions evaluated at the
same expansion scale $\Delta m^2 = 0.3$ GeV$^2$, the CDM boundary
condition of Eq.~(\ref{eq:cdm}), which determines $\sigma_h(Q^2)$
from the transverse-size distribution selected at the production vertex,
attains $\chi^2/N = 0.62$, compared to 3.45 for the conventional
$1/Q^2$ prescription. Since the subsequent QDM transport dynamics are
left unchanged, this improvement isolates the role of the initial PLC
boundary condition. Within the present framework, the CLAS data
therefore support the onset of color transparency in the $\rho^0$
channel beyond what kinematic decay-length effects can accommodate.

The fitted value $\Delta m^2 = 0.3$ GeV$^2$ should be understood as an
effective in-medium expansion scale of the present semi-classical
transport framework rather than a literal $\rho'$ excitation mass gap,
and cannot be directly compared with values obtained in standard QDM
implementations. More broadly, the results indicate that the microscopic
specification of the initial PLC plays a more decisive role in CT
analyses than previously recognized. Future higher-statistics
measurements over an extended $Q^2$ range will be essential for
disentangling the initial configuration selection from the subsequent
in-medium expansion and for testing the onset of CT in the vector-meson
sector with greater stringency.

\begin{acknowledgments}
This work was supported by the National Research Foundation of
Korea (NRF) under Grant No.~NRF-2022R1A2B5B01002307 and by the
Institute for Basic Science (IBS-R031-D1).
\end{acknowledgments}

\section*{Data availability}
The data supporting the findings of this study are available
within the article and from the references cited therein.

\end{document}